\documentclass[11pt]{article}
\usepackage[margin=1in]{geometry}
\usepackage{graphicx}
\usepackage{wrapfig}
\usepackage{authblk}
\usepackage{hyperref}

\def\beq{\begin{equation}}
\def\eeq#1{\label{#1}\end{equation}}
\def\eeqn{\end{equation}}

\def\beqa{\begin{eqnarray}}
\def\eeqa#1{\label{#1}\end{eqnarray}}
\def\eeqan{\end{eqnarray}}

\let\bar=\overbar

\def\Dslash{\not{\hbox{\kern-4pt $D$}}}
\def\dslash{\not{\hbox{\kern-2pt $\del$}}}

\def\msb{{\bar{\ssstyle M \kern -1pt S}}}

\def\Title#1{\begin{center} {\Large {\bf #1} } \end{center}}
\def\Author#1{\begin{center} {\normalsize {\sc #1} } \end{center}}

\def\Abstract#1{\noindent {\normalsize {\bf Abstract:} {\normalfont #1}}}
\def\Conference{\vspace{4mm}\begin{raggedright} {\normalsize {\it Talk presented at the 2019 Meeting of the Division of Particles and Fields of the American Physical Society (DPF2019), July 29--August 2, 2019, Northeastern University, Boston, C1907293.} } \end{raggedright}\vspace{4mm}}

\begin{document}

%
%

\Title{Cosmic Ray isotopes measured by AMS02}

\Author{F.~Dimiccoli$^{1}$,
R.~Battiston$^{1}$,
K.~Kanishchev$^{1}$,
F.~Nozzoli$^{1}$
P.~Zuccon$^{1}$ $^{2}$\\
$^{1}$ INFN, Trento Institute for Fundamental Physics and Applications, I-38123, Trento, Italy\\
$^{2}$ Universit\`a degli Studi di Trento, I-38123 Trento,  Italy}

\Abstract{The measurement of light isotope components in cosmic rays (CR) like $^{3}\mathrm{He}$, $^{2}\mathrm{H}$ (D) and $^{6}\mathrm{Li}$ are some of the most valuable tools for understanding the propagation of CR in the galaxy and constrain the models that describe it.  In this work, new preliminary measurement of $^{3}\mathrm{He}$/$^{4}\mathrm{He}$, $^{6}\mathrm{Li}$/$^{7}\mathrm{Li}$ and D/p flux ratios are presented, each obtained from the data of the AMS-02 experiment, along with the measurement of the isotopic component fluxes.}

\Conference

%
%
\begin{wrapfigure}{R}{6cm}
\centering
\vspace{-1cm}
\includegraphics[width=6cm]{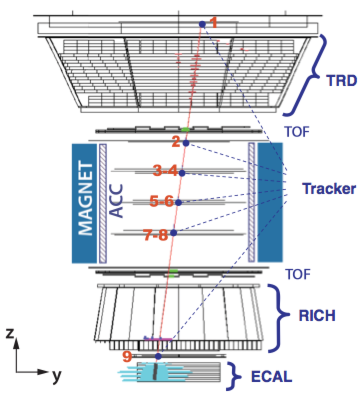}
\caption{\small{\textit{Scheme of the AMS-02 detector. From the top, tracker first layer (1), TRD, upper ToF, inner tracker layers (2,3,4,5,6,7,8), lower ToF, RICH, tracker layer 9 and ECAL. Sixteen curved scintillator panels (Anti-Coincidence Counters, ACC)
surround the inner tracker inside the 0.14T magnet bore.}}\label{fig:ams}}
\end{wrapfigure}
\section{Light isotopic components in CR}

The importance of the measurement of light isotopic components in CR like $^{3}\mathrm{He}$, $^{6}\mathrm{Li}$ and D resides in their production by spallation reactions of primary CR protons and $^{4}\mathrm{He}$ nuclei on the protons of the Inter Stellar Medium (ISM). The measurement of these components with respect to their respective primary can provide important constrains about the CR residence time within galaxy and their propagation history \cite{ref:coste,ref:Tom}. 
Light isotopes originate from the spallation of heavier ones, for example $^{3}\mathrm{He}$ nuclei are produced by the spallation of primary $^{4}\mathrm{He}$ nuclei, and D from $^{3}\mathrm{He}$ and $^{4}\mathrm{He}$. Another source of deuteron in CR is the p-p fusion in the CR collision with the ISM at extremely low energies (below 1 GeV/n). Unlike the above mentioned, both of the $Li$ isotope in CR, $^{6}\mathrm{Li}$ and $^{7}\mathrm{Li}$, are secondary, and come from the fragmentation of heavier species, like $C$, $N$ and $O$.
The relative, secondary  to primary, abundances provide additional and complementary information with respect to the most common tool used for propagation studies, the $B/C$ flux ratio. Given the smaller interaction cross sections the light isotopes probe different propagation distances and, in the deuteron case, the low energy threshold of the p-p fusion reaction makes the D/p ratio measurement useful to constrain propagation models at very low energies\cite{ref:Ama}.

\section{The Instrument: AMS-02}

The Alpha Magnetic Spectrometer (AMS-02) \cite{ref:amssite,ref:ams,ref:ams1,ref:ams2,ref:ams3,ref:ams4,ref:ams5,ref:ams6,ref:ams7,ref:ams8,ref:ams9,ref:ams10,ref:ams11,ref:ams12,ref:ams13,ref:ams14} is a particle detector installed on International Space Station, whose components include a solid state Tracker for determination of charge sign, charge value and rigidity, defined as $R=p/Z$ (where $p$ is the particle momentum), a Time of Flight detector (ToF) and a Cherenkov detector (RICH) for particle velocity, a Transition Radiation Detector (TRD) and an Electromagnetic CALorimeter (ECAL) for hadron/lepton discrimination. The Tracker is composed by 7 layers inside the magnetic volume (inner Tracker), plus two external layers, one above the TRD and one in face of ECAL (respectively, L1 and L9) (fig \ref{fig:ams}).
Thanks to its wide acceptance and privileged position above the Earth's atmosphere, AMS-02 has been providing compelling science, having collected more than 150 billions of charged CR since 2011. Data from published AMS02 results are stored online in the ASI/SSDC cosmic ray database\cite{ref:crdb}.  

\begin{figure}
    \centering
    \includegraphics[height=5cm,width=16cm]{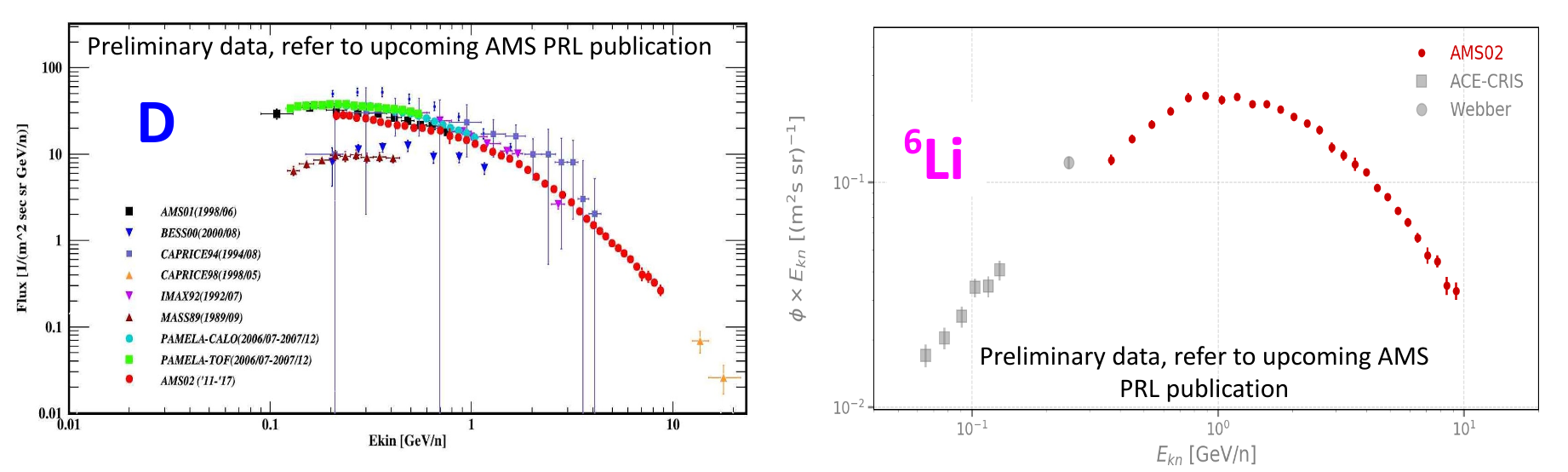}
    \caption{\textit{Deuteron (left) and $^{6}\mathrm{Li}$ (right) fluxes measured as a function of Kinetic energy per nucleon and compared with precedent experiments.}}
    \label{fig:ekinflux}
\end{figure}

\section{Light isotope distinction with AMS-02}
The identification of D, $^{3}\mathrm{He}$ and $^{6}\mathrm{Li}$ isotopes from the respective dominant components (p,$^{4}\mathrm{He}$) is performed in AMS-02 with the concurrent measurement of rigidity and velocity, which when combined provide a mass measurement through the relativistic relation: 
\begin{equation} m=ZR/\gamma\beta. \label{massformula}\end{equation}.
\\This mass separation is performed starting from $Z=1$, $Z=2$ and $Z=3$ samples, obtained with charge identification at different levels within the detector. Additional selection have been implemented to reject most of events suffering of interactions within the detector. 
The rigidity is measured by the inner Tracker, while the velocity by ToF and RICH. The resolution of ToF ($\sim 2\%$) allows isotopic distinction up to 0.85 GeV/n. At higher energies, AMS-02 RICH is used. This detector is equipped with two different radiators, Sodium Fluoride (NaF) and Aerogel (Agl), with different values of thresholds and resolutions, allowing isotopic distinction in the kinetic Energy per nucleon range  0.7 to 3.2 GeV/n and from 2.7 to 9 GeV/n for NaF and Agl respectively. 
\\In the combined range 0.2-9 GeV/n, the rigidity measurement has a resolution of $8 \%$, which dominates the overall mass resolution. Such limited resolutions prevent the event-by-event identification of isotopes and makes a template fit approach on the reconstructed mass distribution necessary for each analysis. 
\subsection{Isotope separation}
The separation is performed dividing each range in narrow bins of measured $\beta$, to exploit the higher precision of the velocity measurement and minimize migration effects, and fitting the measured mass distribution in data with mass templates, obtained for each isotope in each velocity bin.
\begin{figure}
    \centering
    \includegraphics[height=5cm,width=16cm]{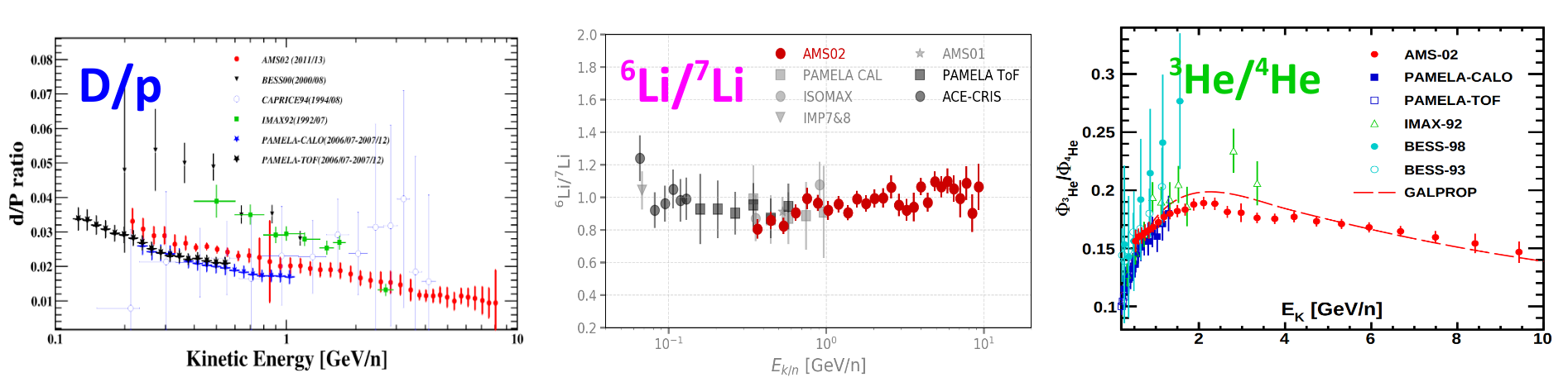}
    \caption{\textit{D/p (left), $^{6}\mathrm{Li}$/$^{7}\mathrm{Li}$ (center) and $^{3}\mathrm{He}$/$^{4}\mathrm{He}$ flux ratios measured as a function of Kinetic energy per nucleon and compared with precedent experiments.}}
    \label{fig:ekinratio}
\end{figure}

The templates for the separation of He and Li isotopes come from analytical functions obtained from the combination of rigidity and velocity response functions studied through the MC simulations. 
For the deuteron case, the templates have been obtained from MC simulations, tuned on both test beam and flight data to reproduce the mass distributions observed in every velocity bin. A selection based on multivariate analysis using Boosted Decision Trees (BDT) was developed to improve the signal to noise ratio in the mass region of interest. 

\subsection{Evaluation of energy loss and fragmentation effects} These two effects have to be carefully evaluated to obtain the correct ratios at the Top Of Instrument (T.O.I.). The energy loss is caused by ionization within in the detector before the inner tracker, mostly in TRD and ToF, and affects the measured energy dependence. The fragmentation regards the production of secondary nuclei of
from the fragmentation of primary ones, mostly due to interactions in the external material of AMS-02. Each of the two effects were evaluated on MC simulations and then validated and cross checked on flight data. 
\begin{figure}
    \centering
    \includegraphics[height=5cm,width=16cm]{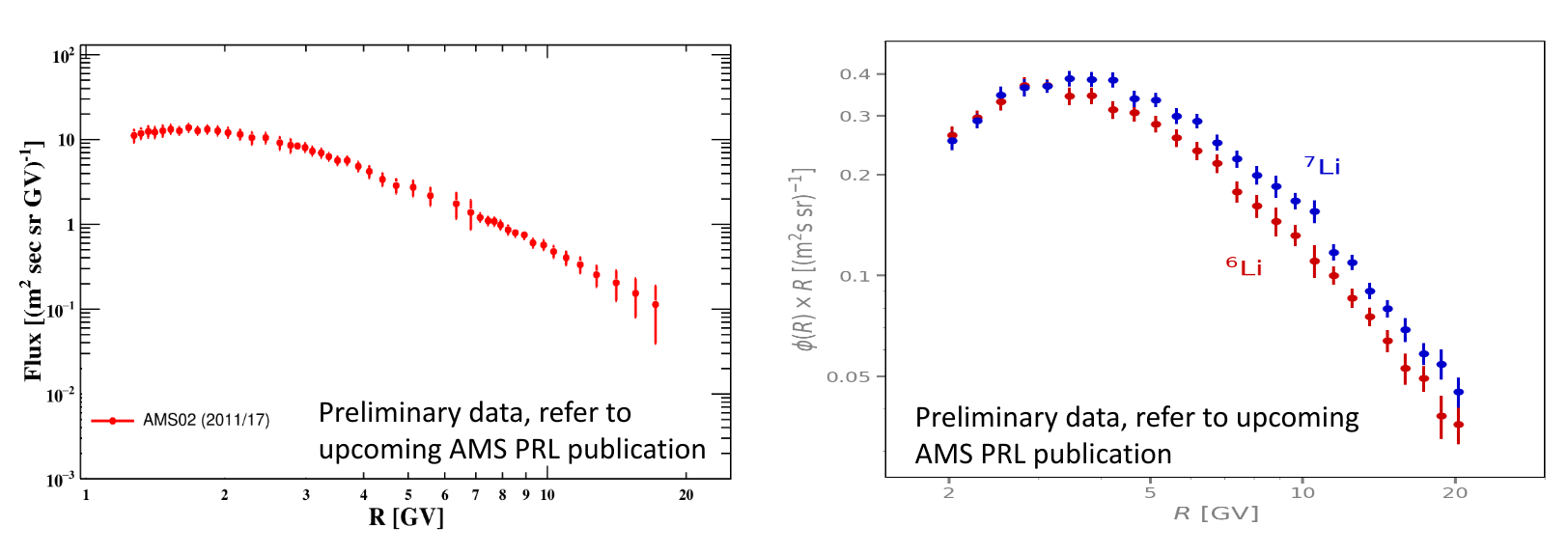}
    \caption{\textit{Deuteron (left) and Lithium (right) fluxes measured as a function of rigidity by AMS-02}}
    \label{fig:rfluxes}
\end{figure}

\section{Isotopic flux and ratio measurements}
Traditionally, isotopic measurements in CR are done as a function of kinetic energy per nucleon, obtained directly from $\beta$ through this relation:\\
$Ekin/n  = (\gamma -1)\frac{mass}{n}$, with $\gamma = \sqrt[]{\frac{1}{1-\beta^{2}}}$, assuming as a constant the average nucleon mass, $\frac{mass}{m}$. Such quantity is thus directly related to the measured velocity.

\begin{figure}
    \centering
    \includegraphics[height=5cm,width=16cm]{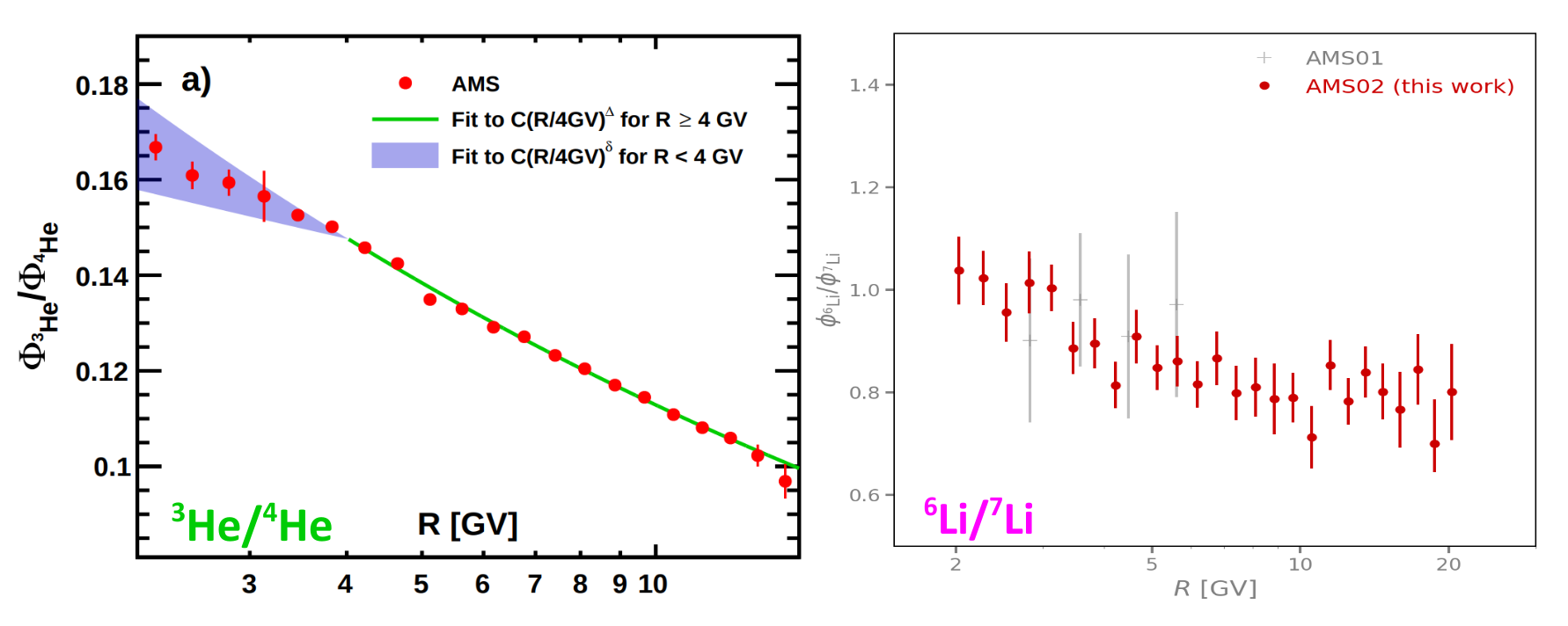}
    \caption{\textit{\textbf{Left:} $^{3}\mathrm{He}$/$^{4}\mathrm{He}$ ratio measured as a function of rigidity. The shadowed area represents the low energy time variability over a period of 6.5 years. \textbf{Right:} $^{6}\mathrm{Li}$/$^{7}\mathrm{Li}$ flux ratio measured as a function of rigidity.}}
    \label{fig:rratio}
\end{figure}
\begin{figure}
    \centering
    \includegraphics[height=10cm,width=16cm]{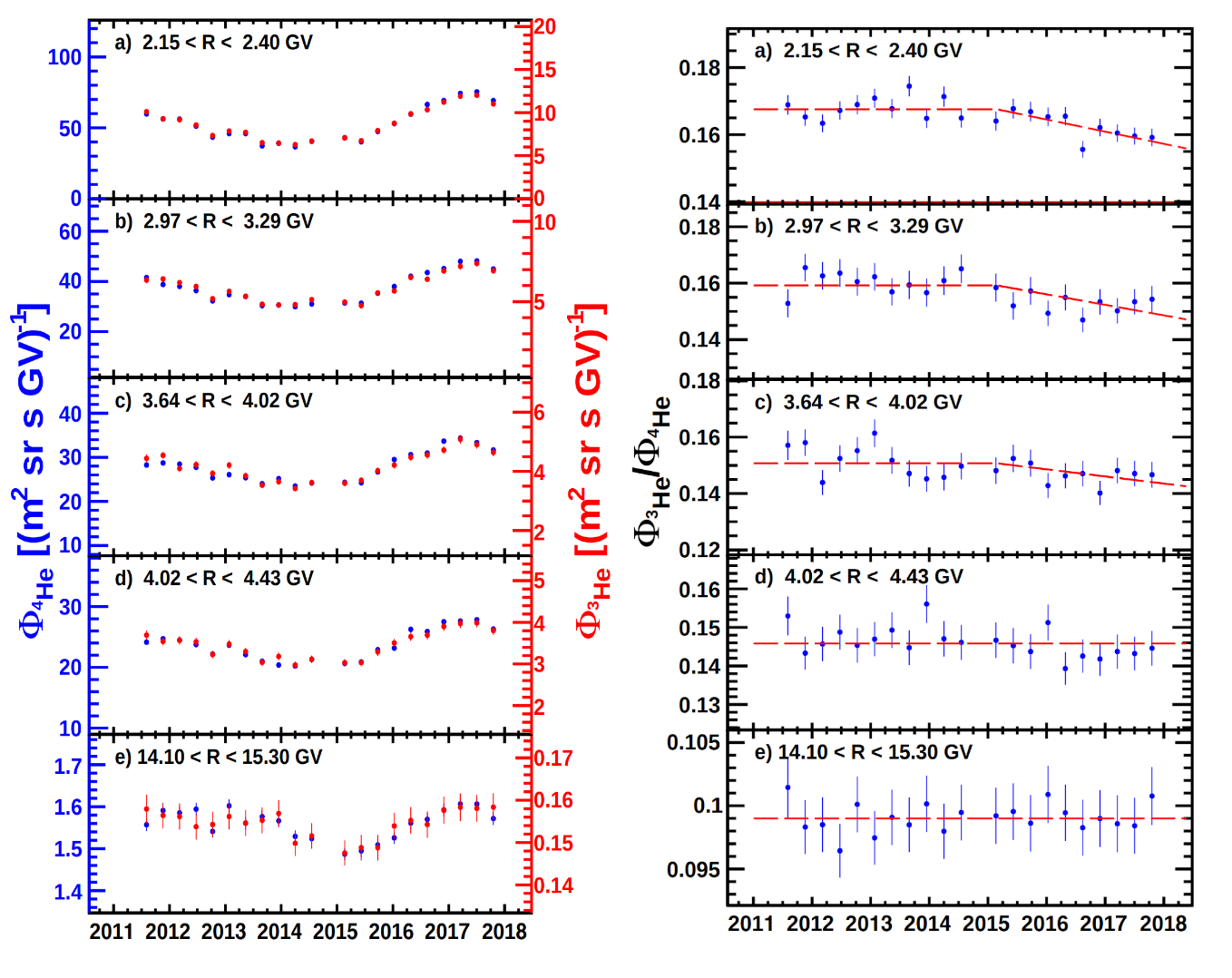}
    \caption{\textit{Time dependence of $^{3}\mathrm{He}$ and $^{4}\mathrm{He}$ fluxes (left) and $^{3}\mathrm{He}$/$^{4}\mathrm{He}$ ratio (right) at different rigidity bins. AMS-02 discovered a low energy time dependence of the ratio, decreasing after Feb $25^{th}$, 2015}}
    \label{fig:timedephe}
\end{figure}

\begin{figure}
    \centering
    \includegraphics[height=7cm,width=16cm]{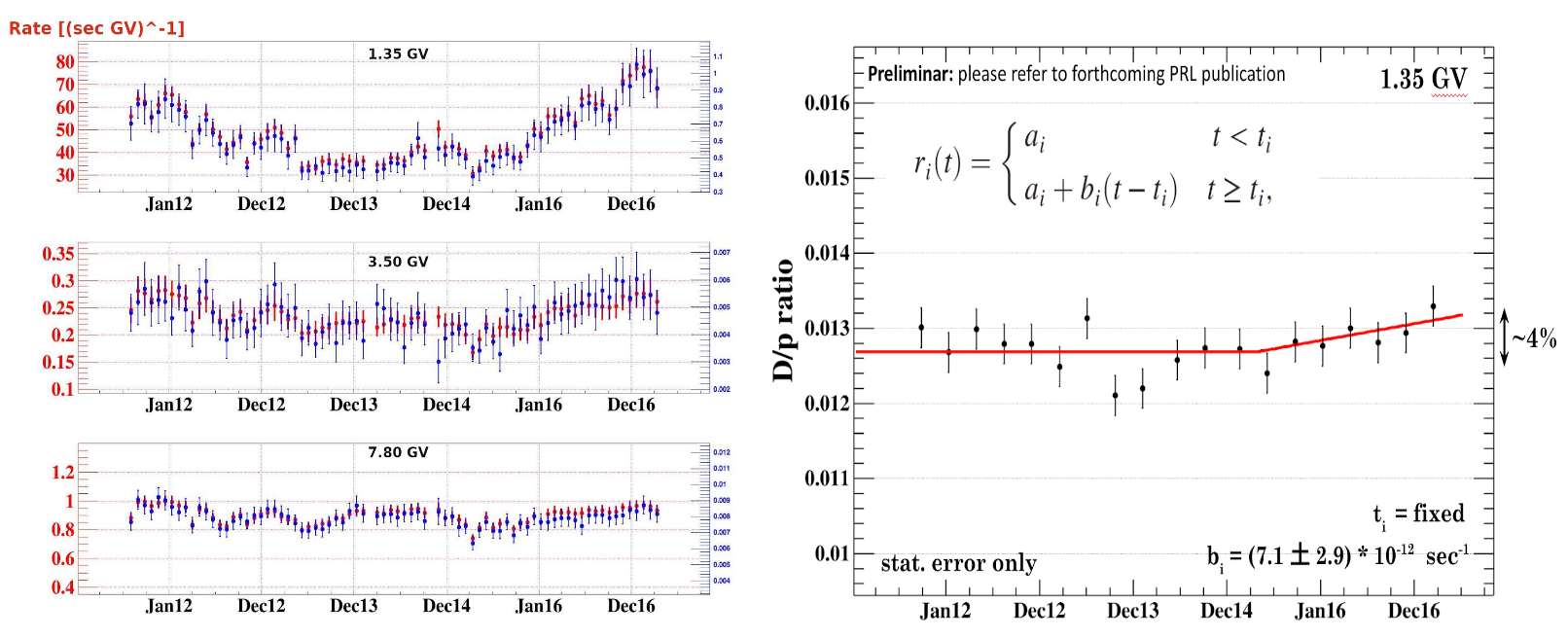}
    \caption{\textit{Time dependence of D and p fluxes at different rigidities (left) and  D/p ratio int the rigidity bin centered around 1.5 GV(right). First hints of a time dependence with the same break at Feb $25^{th}$, 2015 are measured.}}
    \label{fig:timedepd}
\end{figure}

Measuring isotopic fluxes and ratios as a function of kinetic energy per nucleus allows to put the AMS-02 measurements in the framework of precedent measurements. 
AMS-02 can exploit a much wider energy range with respect to its predecessors, exploring substantially uncharted energy regions. In particular, it performed the first measurement of $^{6}\mathrm{Li}$ and $^{7}\mathrm{Li}$ fluxes above 0.3 GeV/n and the first measurement of D flux above 1 GeV/n. This is true also for isotope flux ratio measurements (fig \ref{fig:ekinflux}). AMS02 allowed to obtain the first measurement of $^{3}\mathrm{He}$/$^{4}\mathrm{He}$ ratio above 3 GeV/n and the first measurements of $D/p$ and $^{6}\mathrm{Li}$/$^{7}\mathrm{Li}$ ratios above 1 GeV/n (fig. \ref{fig:ekinratio}).

The high statistic collected by AMS-02 allows to obtain time dependent studies of the isotopic components, and thus a measurement as a function of rigidity was also performed. This results allow to better understand the relation between isotopic fluxes and solar modulation, which is a rigidity dependent effect. 
The measurement is accomplished performing a full isotopic analysis using different velocity bin edges for each isotope, corresponding to a common rigidity binning, taking also into account the effect of energy loss from the top of the instrument.\\
Figure \ref{fig:rfluxes} shows the measurement of D, $^{6}\mathrm{Li}$ and $^{7}\mathrm{Li}$ fluxes, while figure \ref{fig:rratio} shows the measurement of $^{6}\mathrm{Li}$/$^{7}\mathrm{Li}$ ratio.
In general, low energy CR flux is anti-correlated with solar activity. AMS-02 measured CR fluxes over the complete $24^{th}$ solar cycle (from 2011 to 2018), in which period of maximum activity was between 2014 and 2016. The consequent trend on the fluxes of $^{3}\mathrm{He}$ and $^{4}\mathrm{He}$ is visible in the figure \ref{fig:timedephe}, with an effect less and less important at higher energies.
Since the interaction with time dependent solar magnetosphere is rigidity dependent one expects an almost time independent isotope flux ratios as a function of rigidity. Nevertheless, results about $^{3}\mathrm{He}$/$^{4}\mathrm{He}$ time dependent ratio show that actually AMS-02 discovered a small time dependence in the extremely low energy part of the spectrum. The measured trend is constant until a break at Feb $25^{th}$ of 2015, where a linearly descending slope starts. 
First results from D/p time dependent measurement as a function of rigidity seems to give hints for a similar behaviour (fig. \ref{fig:timedepd}).

\section{Conclusions} Isotopic composition of light nuclei in CR is a key measurement to understand cosmic rays origin and propagation. Pursuing this view, AMS-02 has performed the first measurement of $^{6}\mathrm{Li}$ and $^{7}\mathrm{Li}$ fluxes above 0.3 GeV/n and the first precision measurement of D flux above 1 GeV/n. Important tools for the constraint of CR are the measurement of isotopic ratios. In this context, AMS-02 produced measurements of  $^{6}\mathrm{Li}$/$^{7}\mathrm{Li}$, $^{3}\mathrm{He}$/$^{4}\mathrm{He}$ and D/p ratios. Both flux and ratio measurements were performed in terms of kinetic energy per nucleon and rigidity. Finally, AMS-02 discovered a low energy time dependence in the $^{3}\mathrm{He}$ and $^{4}\mathrm{He}$ ratio, and found hints of similar time dependence also in the D/p ratio.

\end{document}